\documentclass[10pt,journal,comsoc]{IEEEtran}

\usepackage{cite}
\usepackage{amsmath,amssymb,amsfonts}
\usepackage[normalem]{ulem}
\usepackage{graphicx}
\usepackage{textcomp}
\usepackage{svg}      
\usepackage{xcolor}
\usepackage[most]{tcolorbox}
\usepackage[table]{xcolor}
\usepackage[caption=false,font=footnotesize]{subfig} 
\usepackage[hidelinks   ]{hyperref}
\usepackage{listings}
\usepackage{multirow}
\usepackage{graphicx}     
\usepackage{caption}      
\tcbset{colback=gray!10!white, colframe=gray!30!black, boxrule=0.4pt, arc=2pt,
        left=6pt,right=6pt,top=6pt,bottom=6pt, fonttitle=\bfseries}
\usepackage{subcaption}   
\usepackage{comment}
\usepackage{pifont}
\usepackage{booktabs}
\usepackage{colortbl}
\usepackage{array}

\tcbset{
  colback=white, colframe=black!35,
  boxsep=1pt, left=4pt, right=4pt, top=3pt, bottom=3pt,
  before skip=4pt, after skip=4pt,
  fonttitle=\bfseries\footnotesize
}
\usepackage{algpseudocode}\usepackage[linesnumbered, ruled]{algorithm2e}
\usepackage[linesnumbered, ruled]{algorithm2e}

\SetKwRepeat{Do}{do}{while}%

\SetCommentSty{mycommfont}

\makeatletter
\newcommand{\algmargin}{\the\ALG@thistlm}
\makeatother

\newlength{\forwidth}
\algdef{SE}[parFOR]{parFor}{EndparFor}[1]
{\parbox[t]{\dimexpr\linewidth-\algmargin}{%
\hangindent\forwidth\strut\algorithmicfor\ #1\ \algorithmicdo\strut}}{\algorithmicend\ \algorithmicfor}%

\newlength{\whilewidth}
\algdef{SE}[parWHILE]{parWhile}{EndparWhile}[1]
{\parbox[t]{\dimexpr\linewidth-\algmargin}{%
\hangindent\whilewidth\strut\algorithmicwhile\ #1\ \algorithmicdo\strut}}{\algorithmicend\ \algorithmicwhile}%
\algnewcommand{\parState}[1]{ %
\parbox[t]{\dimexpr\linewidth-\algmargin}{\strut #1\strut}}

\usepackage{epstopdf}

\begin{document}

\title{Intent2QoS: Language Model-Driven Automation of Traffic Shaping Configurations}
\author{
\IEEEauthorblockN{Sudipta Acharya~and~Burak Kantarci}\\
\thanks{Sudipta Acharya and Burak Kantarci are with University of Ottawa, Ottawa, ON, Canada. Emails: \{sacharya2,burak.kantarci\}@uottawa.ca\\
}
}


\IEEEtitleabstractindextext{
\begin{abstract}
Traffic shaping and Quality of Service (QoS) enforcement are critical for managing bandwidth, latency, and fairness in networks. These tasks often rely on low-level traffic control settings, which require manual setup and technical expertise. This paper presents an automated framework that converts high-level traffic shaping intents in natural or declarative language into valid and correct traffic control rules. To the best of our knowledge, we present the first end-to-end pipeline that ties intent translation in a queuing-theoretic semantic model and, with a rule-based critic, yields deployable Linux traffic control configuration sets. The framework has three steps: (1) a queuing simulation with priority scheduling and Active Queue Management (AQM) builds a semantic model; (2) a language model, using this semantic model and a traffic profile, generates sub-intents and configuration rules; and (3) a rule-based critic checks and adjusts the rules for correctness and policy compliance. We evaluate multiple language models by generating traffic control commands from business intents that comply with relevant standards for traffic control protocols. Experimental results on 100 intents show significant gains, with LLaMA3 reaching 0.88 semantic similarity and 0.87 semantic coverage, outperforming other models by over 30\%. A thorough sensitivity study demonstrates that AQM-guided prompting reduces variability threefold compared to zero-shot baselines. 


\end{abstract}

}
\pagestyle{empty}

\maketitle
\thispagestyle{empty}
\IEEEdisplaynontitleabstractindextext
\IEEEpeerreviewmaketitle

\section{Introduction}\label{intro}


Intent-Based Networking (IBN) has emerged as a promising approach to simplify network management \cite{leivadeas2022survey,clemm2022intent}. In IBN, administrators express high-level goals as “intents”, which the system should interpret and enforce automatically. Recent research demonstrates that Language Models (LMs) are crucial for intent translation, effectively bridging human-level expressions and low-level network configurations \cite{dzeparoska2023llm, towardsE2E}. While progress has been made in validation and orchestration \cite{towardsE2E}, the final step, i.e., translating intents into concrete traffic control (\texttt{tc}) settings often remains manual \cite{wang2024netconfeval,ferguson1998quality}. Administrators still handcraft \texttt{tc} commands by choosing queuing methods and parameters, a process that is error-prone and hard to scale. Despite advances in IBN, little work has focused on automating the translation of high-level intents into low-level QoS enforcement with Linux traffic control.
\begin{figure}[htbp]
  \centering
\includegraphics[width=0.5\textwidth]{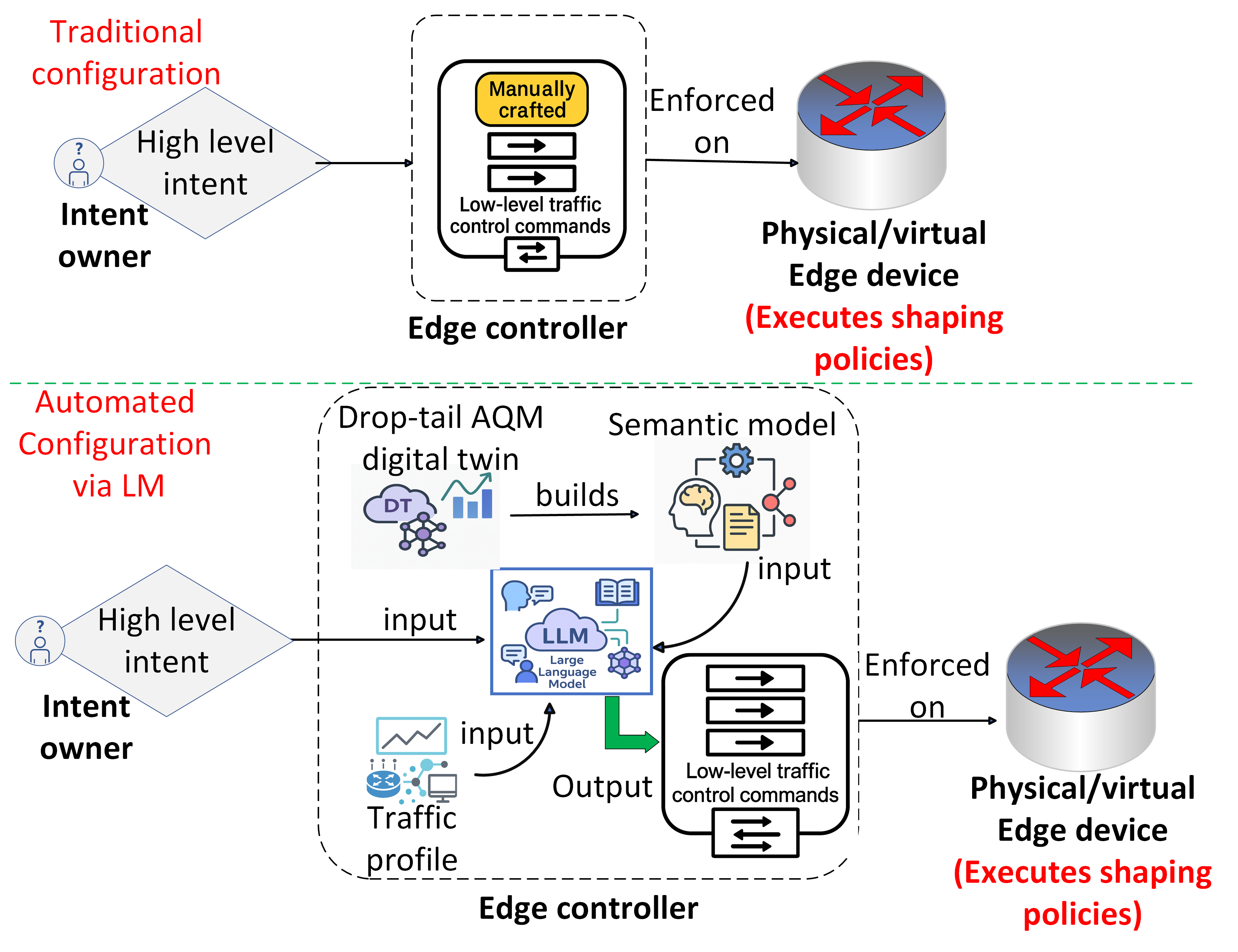}
\caption{Comparison between traditional manual configuration (top) and the proposed automated pipeline (bottom)}
  \label{fig:manual-vs-automated}
\end{figure}

To address the gap between high-level intent and low-level configuration, we introduce an LM-based framework that automates the entire translation pipeline. 
In the manual approach, the intent owner writes the detailed low-level configuration. In our proposed automation solution as shown in Fig.~\ref{fig:manual-vs-automated}, the user provides a high-level intent, and the framework automatically generates valid \texttt{tc} commands ready to be deployed on routers, switches, or edge devices.



Our approach translates high-level network intents into precise Linux \texttt{tc} configurations through a three-phase pipeline. First, we build a \textit{Digital Twin} (DT) of a priority-based queuing system with AQM to capture delay, drop, and utilization behaviors across traffic classes. This forms a \textit{semantic model} that informs intent interpretation. Second, an LM generates sub-intents; a core phase in our framework that extracts and formalizes structured constraints from high-level intents, such as delay bounds, traffic types, and scheduling preferences, and \texttt{tc} configurations aligned with this model. Third, a rule-based critic ensures correctness and policy compliance. 


\noindent In summary, the key contributions of this work are as follows:
\begin{itemize}

\item A semantic model derived from the digital twin of a non-preemptive priority queue with AQM, enabling context-aware interpretation of intents involving delay, drop rate, and priority.
\item A novel framework that translates high-level traffic shaping intents into low-level, platform-compatible \texttt{tc} configurations using LM, derived semantic model of DT, traffic profile and a deterministic post-processing critic module.

\end{itemize}
To the best of our knowledge, there is no prior end-to-end framework that translates high-level network intents into deployable Linux \texttt{tc} configuration sets; consequently, there is no established baseline for a head-to-head comparison. We therefore introduce a 100-intent benchmark and evaluate multiple LMs and prompting regimes within our pipeline. We evaluate our approach against four open-source models: three LLMs (LLaMA3 (8B)~\cite{touvron2024llama3}, Mistral (7B)~\cite{jiang2023mistral}, Gemma (7B)~\cite{deepmind2024gemma}) and one SLM (Phi-2, 2.7B). Our method consistently outperforms these baselines, with LLaMA3 (8B) reaching 0.88 semantic similarity, 0.87 semantic unit coverage, and 0.16 normalized edit distance under AQM-guided two-shot prompting, showing close alignment with reference sub-intents and configurations.


\newcommand{\cmark}{\ding{51}} 
\newcommand{\xmark}{\ding{55}} 

\section{Related Work and The State of the Art}
Traffic shaping and QoS enforcement have long relied on low-level tools such as Linux \texttt{tc} for configuring queuing disciplines, packet classification, and rate limiting. However, despite its maturity, the \texttt{tc} tool remains complex, with limited abstraction and steep learning curves, making automated configuration generation a persistent challenge. In ~\cite{pfefferle2022ats}, Pfefferle et al. explores the implementation of IEEE 802.1Qcr Asynchronous Traffic Shaping using Linux \texttt{tc}, highlighting its manual complexity and limited abstraction. 

To overcome the complexities and limitations of manual configuration, recent research explores automation via data-driven and AI-assisted methods. For example, Wang et al. ~\cite{wang2024netconfeval} analyzed if LLMs like GPT-4 can generate valid network configurations. While the syntax is often correct, the semantic accuracy and policy compliance suffer without system-level awareness. Similarly, Li et al. ~\cite{li2024preconfig} introduced a pre-trained model to assist with network configuration tasks, but it stops short of integrating queue-theoretic reasoning or enforcing low-level QoS constraints. More recent works, such as Dzeparoska et al.~\cite{dzeparoska2024intent} and Lam et al.~\cite{towardsE2E}, attempt to use LLMs for validating or managing intents but do not explore the synthesis of low-level configurations like \texttt{tc}.
As highlighted in the IETF draft by Clemm et al.~\cite{clemm2022intent}, the final step of converting validated intents into deterministic and platform-specific configurations remains an open problem.


 As observed in the existing literature, despite the growing interest in IBN, current solutions offer limited support for translating declarative intents into precise QoS enforcement rules at the system level. This work fills that gap by introducing a novel approach that integrates semantic queue modeling with language model-guided code generation to enable end-to-end automation.

\begin{figure}[!htb]
  \centering
\includegraphics[width=1\linewidth]{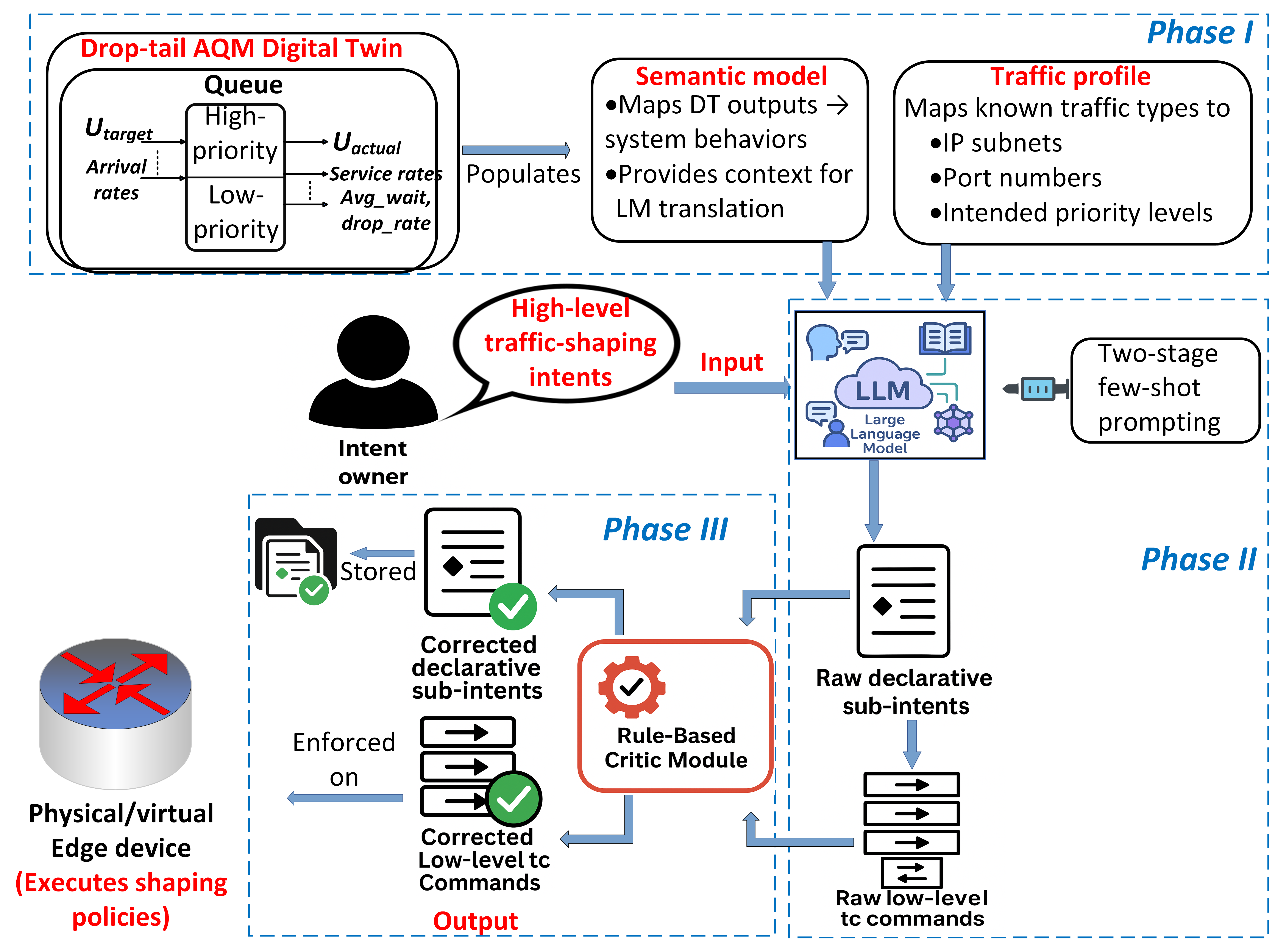}
  \caption{Overview of the proposed edge-executed pipeline for intent-based traffic control}
  \label{fig:intent-pipeline}
\end{figure}

\section{Methodology}
The overview of the proposed intent-to-configuration translation pipeline is provided in Fig.~\ref{fig:intent-pipeline}, and the architecture is explained step by step below. Also, the end-to-end intent-translation process is summarized in Algorithm \ref{alg:intent2tc}.

\paragraph{Input} The input to Phase II is a high-level traffic goal, expressed by the intent owner (e.g., network operator, system administrator, or application developer) in natural language (e.g., latency, bandwidth, priority). System uses this input to initiate sub-intent generation and semantic validation. 


\paragraph{Phase I - Semantic modeling via AQM digital twin and traffic profiling} In the first phase, a \textit{digital twin} of a non-preemptive priority queue with bounded capacity is built using the \texttt{simpy} framework. The model is calibrated offline through queuing-theoretic analysis and traffic modeling to capture realistic behavior under varying loads and priorities. It employs a drop-tail AQM policy, where excess occupancy triggers low-priority packet drops, while remaining extensible to other AQM schemes such as Random Early Detection (RED), Controlled Delay (CoDel), or Proportional Integral controller Enhanced (PIE).

Given inputs such as target utilization ($U_{target}$) and per-class arrival rates, we compute the required service rates $\mu_{\text{high}}$ and $\mu_{\text{low}}$. The model then derives metrics including average waiting times, per-class drop rates, and actual utilization ($U_{actual}$). These metrics populate a \textit{semantic model} that links high-level intent objectives (e.g., minimizing delay) with feasible system behaviors, while capturing trade-offs such as fairness and efficiency. For example, given the intent \textit{“Minimize delay for voice traffic during 8 PM–1 AM”,} the model estimates achievable thresholds under the AQM policy and evaluates the impact on other traffic classes.

To avoid application-layer inspection, a \textit{traffic profile} module maps known traffic types to IP subnets, port ranges, and intended priorities\footnote{Priority levels from the traffic profile are used only when the intent does not explicitly specify them.}. A synthetic profile is constructed with representative traffic classes, using RFC~1918 private address blocks~\cite{rfc1918} and conventional port assignments (e.g., Modbus/TCP on 502~\cite{ianaPorts}, telemetry on 8000–8100, IoT/fire alerts on 30000–50000~\cite{iotSurvey}). This profile is later used in \textit{sub-intent generation}.


\paragraph{Phase II - Intent Decomposition and Translation}
In the second phase, the framework applies a two-stage transformation of high-level natural language intents into executable Linux \texttt{tc} commands, using a language model pipeline enriched with semantic and traffic context from Phase I. 
\\
\emph{Phase II-A - Declarative Sub-Intent Generation:}
In this phase, the LM derives structured sub-intents from the high-level user intent, enriched with traffic attributes from the semantic model and traffic profile. These sub-intents are \emph{declarative} representations of target behaviors (e.g., delay bounds, scheduling priorities) for specific traffic classes. Each of them captures an actionable requirement consistent with system capabilities and queue behavior, serving as an intermediate layer between high-level goals and low-level enforcement logic.

Given a high-level user-specified traffic shaping intent (e.g., \emph{``Prioritize industrial robotics traffic under congestion''}), the system constructs a structured LM prompt by injecting two critical sources of context, i.e., \emph{Semantic model} and \emph{traffic profile}.
The prompt follows a zero, one, or two-shot format as described in Table \ref{tab:prompting-strategies}, illustrating how intents are to be translated into structured declarative sub-intents and then into low-level \texttt{tc} configurations. These resultant sub-intents are designed to be:

\begin{itemize}
    \item \textbf{Quantifiable and enforceable:} e.g., \texttt{avg\_wait\_high $\leq$ 0.13s}\footnote{Average waiting time for high-priority traffic in the queue} or \texttt{drop\_rate\_low $\leq$ 5\%}\footnote{Packet drop rate for low-priority traffic}.
    \item \textbf{Priority-aware:} capturing class hierarchy, e.g., \texttt{assign\_priority(telemetry, low)}.
    \item \textbf{AQM-aligned:} enforcing thresholds that prevent congestion and stabilize latency.
\end{itemize}

The generated sub-intents by LM form a bridge between user-level goals and enforceable \texttt{tc} configurations, which are generated during the next sub-phase. 
\begin{table}[ht]
\centering
\renewcommand{\arraystretch}{1.05}
\setlength{\tabcolsep}{3pt}
\begin{tabular}{|p{1.5cm}|p{2.8cm}|p{2.8cm}|}
\hline
\textbf{Strategy} & \textbf{Intent $\rightarrow$ Sub-intent} & \textbf{Sub-intent $\rightarrow$ Config.} \\
\hline
Zero-shot & No examples; only high-level intent. & No examples; only sub-intents. \\
\hline
One-shot & One intent with sub-intents injected. & One sub-intent with config injected. \\
\hline
Two-shot+AQM & Two examples + semantic model (delay, drop, priority) + traffic profile. & Two examples + same semantic/traffic context. \\
\hline
\end{tabular}
\caption{Prompting strategies for the two translation stages.}
\label{tab:prompting-strategies}
\end{table}

\emph{Phase II-B - Low-Level \texttt{tc} Configuration Generation:} In the second sub-phase, the declarative sub-intents produced in Phase II-A are used as input to a second LM prompt, which synthesizes executable Linux \texttt{tc} configurations that remain consistent with the actual traffic characteristics observed at the network layer. This prompt also includes the relevant traffic profile and semantic model to guide rule enforcement accurately. The language model is instructed to generate valid, platform-compatible \texttt{tc} commands satisfying the following constraints as illustrated in Table \ref{tab:tc_constraints_examples}. 
\begin{table}[ht]
\centering
\resizebox{\columnwidth}{!}{
\begin{tabular}{|p{3.3cm}|p{6.5cm}|}
\hline
\textbf{Constraint Type} & \textbf{Illustrative Example / Explanation} \\
\hline
Class-based queuing hierarchy & Use \texttt{htb} to define traffic classes: e.g., classid \texttt{1:1}, \texttt{1:2} for high and low priority traffic. \\
\hline
Priority mapping & Assign \texttt{prio} values like \texttt{prio 0} (high) and \texttt{prio 2} (low) based on sub-intent directives. \\
\hline
Delay and loss modeling &
Use \texttt{netem delay 120ms loss 2\%} to reflect constraints like
\texttt{avg\_wait\_low} $\leq 0.12\,\text{s}$ or
\texttt{drop\_rate\_high} $\leq 2\%$. \\
\hline
Packet classification & Use \texttt{u32 match ip src 10.1.1.0/24 match ip dport 502 0xffff} to classify industrial robotics traffic. \\
\hline
Temporal annotations & Include comment: \texttt{\# enforce from 20:00 to 01:00} (for intents with time-bound policies). \\
\hline
\end{tabular}
}
\caption{Constraints enforced in \texttt{tc} configuration generation with illustrative examples}
\label{tab:tc_constraints_examples}
\end{table}


The sub-intents and \texttt{tc} configurations generated in this phase serve as raw outputs. They are forwarded to the rule-based critic in Phase III for validation and refinement, ensuring completeness, correctness, and alignment with platform requirements before final deployment.

\paragraph{Phase III: Rule-Based Critic for Post-Validation}

In the final stage, we employ a deterministic rule-based critic that validates and corrects the raw outputs from Phase~II in two steps. First, it examines the raw sub-intents, producing a set of corrected sub-intents that are semantically consistent and complete. Next, it cross-checks the corrected sub-intents against the raw \texttt{tc} configurations, repairing structural gaps and enforcing compliance with kernel-level requirements. The result is a set of corrected \texttt{tc} commands ready for deployment on the physical or virtual edge device.

The corrected sub-intents themselves are retained as a verifiable semantic layer: they enable traceability, auditing, and future re-synthesis without being directly enforced. This two-step process ensures that final configurations are both semantically correct and syntactically executable. 

\paragraph{Output}
The final stage produces a validated configuration package, ready for deployment on Linux edge devices or for downstream evaluation. By resolving vague or under-specified intents, it strengthens the robustness of the pipeline. A case study in the next section illustrates this process.

\begin{algorithm}[!t]
\caption{Intent-to-\texttt{tc} Configuration via Semantic Modeling and Critic Validation}
\label{alg:intent2tc}
\fontsize{7.3}{7.5}\selectfont
\KwData{Natural-language intents $\mathcal{I}$}
\KwResult{Corrected sub-intents $\hat{I}$, valid \texttt{tc} configs $T_{\hat{I}}$}

\textbf{/* Phase I: Semantic Modeling (Queue DT) */}\\
Set $(\lambda_{\text{high}},\lambda_{\text{low}},\mu_{\text{high}},\mu_{\text{low}})$; 
$S \leftarrow$ SimPy priority-queue with AQM; extract $(avg\_wait,drop\_rate,util)$\;

\textbf{/* Phase III (setup): Load critic rules */}\;
Load sub-intent rules $R_d$ and tc rules $R_c$\;

\ForEach{$i \in \mathcal{I}$}{
  \textbf{/* Phase II-A: Sub-intent Generation */}\\
  $K \leftarrow \text{keywords}(i)$;\quad $P_K \leftarrow \text{profile\_filter}(K)$\;
  \emph{build\_prompt}$\,(i,S,P_K)$; \quad
  $I_d \leftarrow \text{LM\_subintents}(i,S,P_K)$\;

  \textbf{/* Phase II-B: \texttt{tc} Configuration Generation */}\\
  \emph{build\_cfg\_prompt}$\,(I_d,S,P_K)$;\quad
  $T_p \leftarrow \text{LM\_tc}(I_d,S,P_K)$\;

  \textbf{/* Phase III: Rule-Based Critic Validation */}\\
  $\hat{I} \leftarrow \hat{I} \cup \text{fix\_subs}(I_d,R_d)$\;
  $T_{\hat{I}} \leftarrow T_{\hat{I}} \cup \text{fix\_tc}(T_p,R_c,\hat{I})$\;
}
\textbf{/* Output */}\;
Save $\hat{I},T_{\hat{I}}$ as JSON; report correction counts and rule violations\;
\end{algorithm}

\begin{figure}[t]
  \centering
  \subfloat[QoS objectives]{%
    \includegraphics[width=0.50\linewidth]{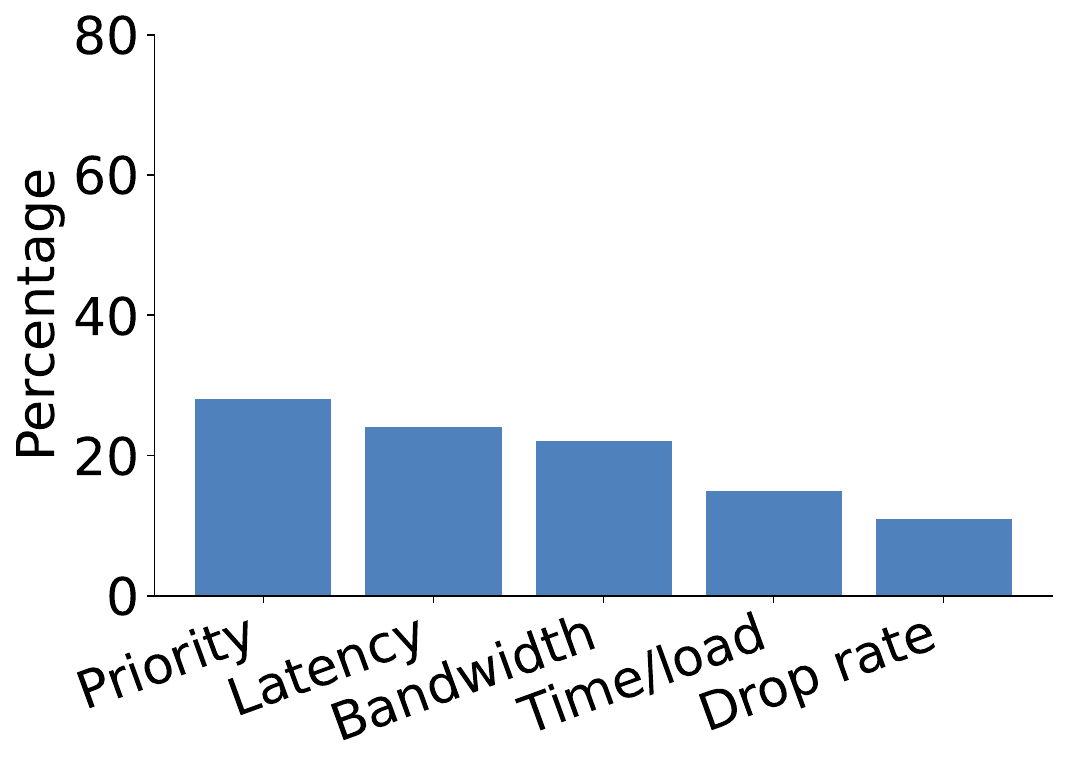}%
    \label{fig:qos}}
  \hfill
  \subfloat[Traffic types referenced]{%
    \includegraphics[width=0.50\linewidth]{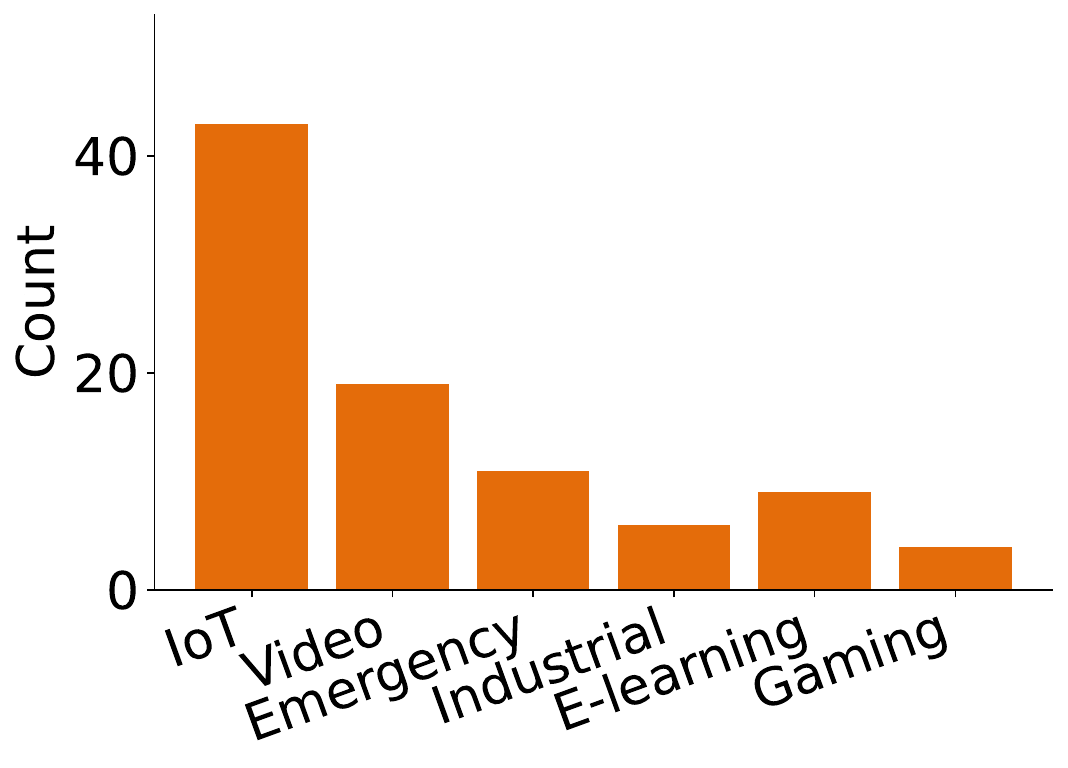}%
    \label{fig:traffic}}

  \vspace{0.05em}
  \subfloat[Intent structure]{%
    \includegraphics[width=0.52\linewidth]{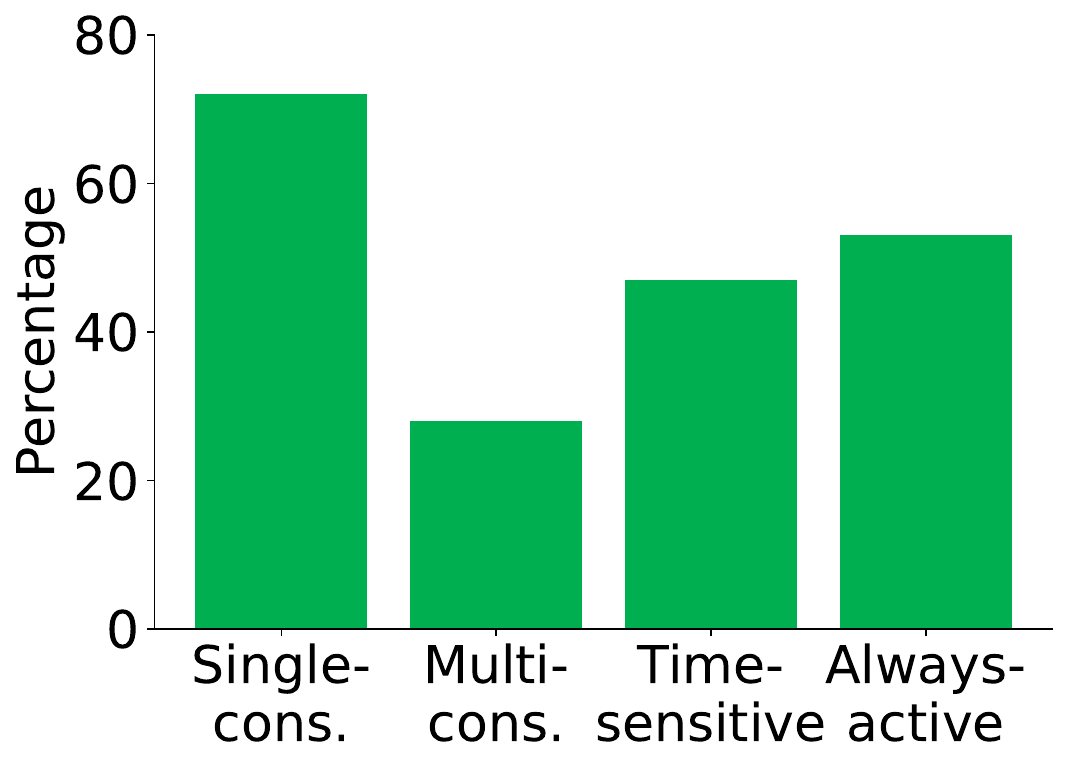}%
    \label{fig:intent}}

  \caption{Summary statistics for the 100-intent dataset}
  \label{fig:dataset-stats}
\end{figure}
\section{Performance evaluation}
\subsection{Dataset}
To evaluate the proposed pipeline, we curated a synthetic dataset of 100 handcrafted high-level network traffic shaping intents. Each intent encodes a distinct user-level policy objective such as latency reduction, drop minimization, bandwidth guarantees, priority assignment, congestion control, or time-sensitive shaping reflecting realistic goals in programmable networks spanning edge-cloud systems, SD-WANs, and ISP cores. The dataset strictly follows the RFC 9315~\cite{clemm2022intent} guidelines, which require intents to specify only the goal i.e., \emph{what} outcome is desired without describing \emph{how} to achieve it. Accordingly, our dataset ensures:  
\begin{itemize}
    \item \textbf{Clarity and Declarativeness:} Natural-language intents define objectives (e.g., latency thresholds, prioritization, rate control) without implementation details.  
    \item \textbf{Abstraction from Device Logic:} No low-level configuration elements (e.g., interface names, command syntax, queue identifiers), ensuring platform neutrality.  
    \item \textbf{Traffic Shaping Semantics:} Each intent captures shaping policies relevant to modern QoS - delay bounds, drop rates, bandwidth caps, and fairness across classes.  
\end{itemize}  

The use of a handcrafted dataset is deliberate and necessary, as no benchmark corpora exist for intent-to-sub-intent or intent-to-configuration translation. Each intent was carefully designed and reviewed to maximize structural diversity, semantic coverage, and realism.

\textbf{Statistical Summary:} 
Fig.~\ref{fig:dataset-stats} summarizes the 100-intent dataset. 
The distribution shows a balanced mix of objectives: priority or fairness (28\%), latency (24\%), bandwidth (22\%), 
time/load-aware (15\%), and drop rate (11\%). 
Most intents specify a single constraint (72\%), while 28\% combine multiple objectives; 
47\% are time-sensitive and 53\% always active. 
The dataset also spans diverse traffic types, including IoT (43 cases), video (19), emergency (11), 
industrial (6), e-learning (9), and gaming (4), reflecting realistic and varied network control scenarios.

The reference sub-intents and \texttt{tc} configurations used for evaluation were curated and validated by domain experts in network traffic engineering. This ensured that the ground truth captures realistic, technically correct enforcement policies against which the generated outputs could be compared. 




\begin{figure*}[t]
  \centering
  \includegraphics[width=0.9\linewidth]{figs/casestudy3.png}
  \caption{End-to-end case study flow: High-level Intent $\rightarrow$ Raw LM outputs (pre-critic) $\rightarrow$ Critic-corrected, deployable configuration
}
  \caption*{\footnotesize\emph{Note:} The \texttt{netem delay} parameter shown in the figure represents an average queuing delay (approximately 142 ms) derived from buffer sizing (e.g., \texttt{bfifo limit}) and HTB scheduling. In practice, delay arises from queue dynamics; future versions will adopt queue-length–adaptive AQM mechanisms (e.g., \texttt{fq\_codel}) to more accurately reflect M/M/1 delay thresholds.}
  \label{fig:case-study-voice}
\end{figure*}

\begin{table*}[t]
\centering
\small
\resizebox{\textwidth}{!}{%
\begin{tabular}{@{}llp{0.46\linewidth}p{0.34\linewidth}@{}}
\toprule
\textbf{Stage} & \textbf{Metric} & \textbf{Formula} & \textbf{Interpretation} \\
\midrule
\multirow{3}{*}{Sub-intent}
  & \textbf{Sentence-BERT (SBERT) similarity}~\cite{reimers2019sentence} \(\uparrow\)
  & \(\displaystyle \frac{\mathbf{e}_g \!\cdot\! \mathbf{e}_r}{\lVert \mathbf{e}_g \rVert \, \lVert \mathbf{e}_r \rVert}\)
  & Semantic similarity (cosine on sentence embeddings). \\
  & \textbf{ROUGE-L F1}~\cite{lin2004rouge} \(\uparrow\)
  & \(\displaystyle \text{F1}=\frac{(1+\beta^{2})\,(\mathrm{LCS\!-\!P}) \cdot (\mathrm{LCS\!-\!R})}{\mathrm{LCS\!-\!P}+\beta^{2}\mathrm{LCS\!-\!R}},\ \beta=1\)
  & LCS overlap (balanced precision/recall). \\
  & \textbf{Token Precision / Recall / F1}~\cite{schutze2008introduction} \(\uparrow\)
  & \(\displaystyle P=\frac{|T^{(\text{sub})}_g \cap T^{(\text{sub})}_r|}{|T^{(\text{sub})}_g|},\ 
               R=\frac{|T^{(\text{sub})}_g \cap T^{(\text{sub})}_r|}{|T^{(\text{sub})}_r|},\ 
               F1=\frac{2PR}{P+R}\)
  & Token-level match vs. reference sub-intents. \\
\midrule
\multirow{3}{*}{\texttt{tc} Gen.}
  & \textbf{Semantic-Unit Coverage}~\cite{peng2020few} \(\uparrow\)
  & \(\displaystyle \mathrm{Cov}=\frac{|U_g \cap U_r|}{|U_r|}\)
  & Required semantic units present (selectors, thresholds, classes). \\
  & \textbf{Token Precision / Recall / F1}~\cite{schutze2008introduction} \(\uparrow\)
  & \(\displaystyle P=\frac{|T^{(\texttt{tc})}_g \cap T^{(\texttt{tc})}_r|}{|T^{(\texttt{tc})}_g|},\ 
               R=\frac{|T^{(\texttt{tc})}_g \cap T^{(\texttt{tc})}_r|}{|T^{(\texttt{tc})}_r|},\ 
               F1=\frac{2PR}{P+R}\)
  & Token-level match for configuration tokens. \\
  & \textbf{Normalized Edit Distance (NED)}~\cite{yujian2007normalized} \(\downarrow\)
  & \(\displaystyle \mathrm{NED}=\frac{\mathrm{Lev}(C_g,C_r)}{\max(|C_g|,|C_r|)}\)
  & Normalized edit distance between generated and reference configs. \\
\bottomrule
\end{tabular}}
\caption{Evaluation metrics for the two translation stages. \textbf{Legend:} \(\uparrow\/ \downarrow\)=higher/lower is better; ROUGE-L F1=ROUGE longest-common-subsequence F1; LCS-P/R=LCS precision/recall; Lev=Levenshtein distance; 
\(T^{(\text{sub})}\)=token set for sub-intents, \(T^{(\texttt{tc})}\)=token set for configurations; \(U\)=semantic-unit set; \(C\)=token \emph{sequence}; \(\mathbf{e}\)=SBERT embedding; \(g,r\)=generated/reference.}
\label{tab:eval-metrics}
\end{table*}

 
\subsection{Case Study: Voice Traffic Delay Minimization}
\label{sec:case-study}
To demonstrate the end-to-end operation of the proposed framework, we present a case study derived from a high-level intent focused on minimizing delay for real-time voice traffic. Fig. \ref{fig:case-study-voice} illustrates the complete workflow - from the initial natural-language intent to the critic-validated, deployable configuration artifacts. The rule-based critic applies the following structural and semantic corrections to the raw Linux \texttt{tc} configuration set, as illustrated in Fig.~\ref{fig:case-study-voice}, prior to deployment:
\begin{itemize}
  \item Removed invalid \texttt{prio} arguments from HTB \texttt{class} definitions.
  \item Normalized thresholds/units to semantic model values (0.142\,s, 0.4\%, 2.8\%).
  \item Fixed filters: added explicit UDP match, corrected RTP masks, removed redundant catch-all.
\end{itemize}

\begin{table}[t]
\centering
\renewcommand{\arraystretch}{1.15}
\setlength{\tabcolsep}{3pt}
\begin{tabular}{|p{1.5cm}|p{0.9cm}|c|c|c|c|c|}
\hline
\textbf{Model} & \textbf{Prompt} & \textbf{Sem.} & \textbf{ROUGE-L} & \textbf{Tok.} & \textbf{Tok.} & \textbf{Tok.} \\
&  &  \textbf{Sim.} & \textbf{F1} & \textbf{Prec.} & \textbf{Rec.} & \textbf{F1} \\
\hline
\multirow{3}{*}{LLaMA3(8B)} 
    & (1)        & 0.39 $\pm$ 0.07 & 0.35 & 0.62 & 0.60 & 0.62 \\
    & (2)        & 0.60 $\pm$ 0.06 & 0.63 & 0.75 & 0.72 & 0.74 \\
    \rowcolor{yellow!20}
    & \textbf{(3)} & \fbox{0.88 $\pm$ 0.03} & \fbox{0.82} & \fbox{0.86} & \fbox{0.85} & \fbox{0.86} \\
\hline
\multirow{3}{*}{Mistral(7B)} 
    & (1)        & 0.35 $\pm$ 0.09 & 0.33 & 0.58 & 0.55 & 0.57 \\
    & (2)        & 0.65 $\pm$ 0.09 & 0.51 & 0.69 & 0.67 & 0.68 \\
    & (3)        & 0.85 $\pm$ 0.07 & 0.78 & 0.83 & 0.80 & 0.82 \\
\hline
\multirow{3}{*}{Gemma(7B)} 
    & (1)        & 0.33 $\pm$ 0.1 & 0.30 & 0.56 & 0.53 & 0.53 \\
    & (2)        & 0.72 $\pm$ 0.09 & 0.50 & 0.67 & 0.65 & 0.66 \\
    & (3)        & 0.81 $\pm$ 0.07 & 0.69 & 0.80 & 0.78 & 0.79 \\
\hline
\multirow{3}{*}{Phi-2} 
    & (1)        & 0.27 $\pm$ 0.12 & 0.33 & 0.50 & 0.48 & 0.48 \\
    & (2)        & 0.59 $\pm$ 0.1  & 0.46 & 0.59 & 0.55 & 0.56 \\
    & (3)        & 0.64 $\pm$ 0.09 & 0.61 & 0.70 & 0.64 & 0.66 \\
\hline
\end{tabular}
\caption{Performance of LMs on Intent-to-Sub-intent Translation. Reported Semantic similarity is mean $\pm$ SD over 10 runs; other metrics are means. Prompts: (1) Zero-shot, (2) One-shot, (3) AQM+Two-shot}
\label{tab:intent_subintent_metrics}
\end{table}


\begin{table}[t]
\centering
\renewcommand{\arraystretch}{1.15}
\setlength{\tabcolsep}{3pt}
\begin{tabular}{|p{1.5cm}|p{0.9cm}|c|c|c|c|c|}
\hline
\textbf{Model} & \textbf{Prompt} & \textbf{Sem. Unit} & \textbf{Tok.} & \textbf{Tok.} & \textbf{Tok.} & \textbf{Norm.} \\
& & \textbf{Cov.} & \textbf{Prec.} & \textbf{Rec.} & \textbf{F1} & \textbf{Edit Dist.} \\
\hline
\multirow{3}{*}{LLaMA3(8B)} 
    & (1)        & 0.43 $\pm$ 0.08 & 0.62 & 0.57 & 0.60 & 0.45 \\
    & (2)        & 0.66 $\pm$ 0.06 & 0.77 & 0.73 & 0.74 & 0.32 \\
    \rowcolor{yellow!20}
    & \textbf{(3)} & \fbox{0.87 $\pm$ 0.04} & \fbox{0.88} & \fbox{0.85} & \fbox{0.86} & \fbox{0.16} \\
\hline
\multirow{3}{*}{Mistral(7B)} 
    & (1)        & 0.43 $\pm$ 0.1 & 0.59 & 0.55 & 0.57 & 0.47 \\
    & (2)        & 0.60 $\pm$ 0.08 & 0.74 & 0.69 & 0.71 & 0.35 \\
    & (3)        & 0.83 $\pm$ 0.04 & 0.84 & 0.83 & 0.84 & 0.30 \\
\hline
\multirow{3}{*}{Gemma(7B)} 
    & (1)        & 0.42 $\pm$ 0.1 & 0.58 & 0.52 & 0.54 & 0.49 \\
    & (2)        & 0.60 $\pm$ 0.1 & 0.70 & 0.68 & 0.68 & 0.35 \\
    & (3)        & 0.75 $\pm$ 0.06 & 0.82 & 0.80 & 0.81 & 0.31 \\
\hline
\multirow{3}{*}{Phi-2} 
    & (1)        & 0.38 $\pm$ 0.13 & 0.49 & 0.45 & 0.47 & 0.56 \\
    & (2)        & 0.57 $\pm$ 0.09 & 0.62 & 0.60 & 0.60 & 0.41 \\
    & (3)        & 0.72 $\pm$ 0.07 & 0.74 & 0.67 & 0.71 & 0.35 \\
\hline
\end{tabular}
\caption{Performance of LMs on Sub-intent-to-Configuration Translation. Reported Semantic Unit Coverage is mean $\pm$ SD over 10 runs; other metrics are means. Prompts: (1) Zero-shot, (2) One-shot, (3) AQM+Two-shot}
\label{tab:intent_tc_metrics}
\end{table}

\subsection{Analysis of obtained results}
We evaluate the pipeline across two translation stages: (i) sub-intent generation from high-level intents, and (ii) translation of sub-intents into low-level \texttt{tc} commands. Table \ref{tab:eval-metrics} describes the chosen metrics in both translation stages to capture semantic fidelity, lexical overlap, structural alignment, and syntactic correctness. 

As noted in Section~\ref{intro}, the absence of baselines motivates a 100-intent benchmark to evaluate multiple LMs end-to-end. We compare generated sub-intents and \texttt{tc} configurations against references using the chosen evaluation metrics, emphasizing (i) semantic fidelity and (ii) structural/functional correctness. We apply zero-shot, one-shot, and AQM-grounded two-shot prompting at both translation stages (Table~\ref{tab:prompting-strategies}). Tables~\ref{tab:intent_subintent_metrics} and~\ref{tab:intent_tc_metrics} report mean performance over 10 runs. Please note that larger models such as LLaMA\,3.2\,(70B) and LLaMA\,4\,(109B total, 17B active) were excluded to maintain fair comparison across dense models ($\leq$8B) and ensure practical inference time for large-scale evaluations.

\subsubsection{Analysis of Generated Sub-intents}
As shown in Table \ref{tab:intent_subintent_metrics}, under the zero-shot setup, all LM models produce syntactically valid outputs but omit essential constraints like priority, delay, and queue behavior. This leads to low semantic similarity, particularly for \textbf{Phi-2} (SBERT score 0.27), which struggles due to limited capacity and lack of context. \textbf{One-shot prompting} improves structural alignment by providing a single example. However, in the absence of semantic grounding, models still hallucinate parameters or misinterpret ambiguous intents. The \textbf{two-shot + AQM} strategy delivers the best performance across all metrics. Semantic model and traffic profile injection provide contextual knowledge such as delay bounds and drop thresholds, enabling more accurate sub-intent generation. \textbf{LLaMA3} achieves top scores (SBERT: 0.88, ROUGE-L:0.82, F1:0.86), with \textbf{Mistral} following closely. \textbf{Phi-2} shows limited improvement, indicating difficulty in leveraging multi-shot context effectively.

Overall, the results confirm that without semantic grounding via AQM, models often miss essential constraints; with it, sub-intents become both meaningful and technically valid - crucial for downstream configuration generation.

\subsubsection{Analysis of Generated \texttt{tc} Configurations}
Based on reported results in Table~\ref{tab:intent_tc_metrics}, we have also evaluated how the selected language models translate declarative sub-intents into Linux \texttt{tc} configurations across three prompting strategies.

In the \textbf{zero-shot} setting, models often omit key semantic units (e.g., \texttt{rate}, \texttt{prio}, \texttt{handle}) and produce structurally flawed commands. This results in low semantic coverage (e.g., 0.43 for LLaMA3) and high NED (e.g., 0.45 for LLaMA3). \textbf{Phi-2}, the smallest model, performs the worst, showing limited contextual understanding without examples. \textbf{One-shot prompting} offers moderate improvement, reducing syntax errors and increasing token-level F1 scores. However, without semantic grounding, models still hallucinate parameters or misinterpret intent semantics, especially in complex or ambiguous cases. The \textbf{two-shot + AQM} strategy yields the best results across all models. \textbf{LLaMA3} leads with 0.87 semantic unit coverage and the lowest NED (0.16), followed by \textbf{Mistral} and \textbf{Gemma}. \textbf{Phi-2} improves slightly, but struggles to leverage multi-shot context effectively. The quality of generated \texttt{tc} configurations is highly dependent on the accuracy of the sub-intents. Errors or omissions in sub-intents propagate into the final commands, emphasizing the need for semantically grounded intermediate representations.

In summary, the combination of few-shot prompting and AQM-based semantic context significantly enhances both the structural validity and functional correctness of generated \texttt{tc} rules.

\subsubsection{Sensitivity Analysis}
We evaluated robustness across all LMs by conducting a sensitivity study at both translation stages: intent to sub-intent and sub-intent to \texttt{tc} configuration. For Semantic Similarity (Table~\ref{tab:intent_subintent_metrics}) and Semantic Unit Coverage
(Table~\ref{tab:intent_tc_metrics}), we report mean scores with standard deviation (SD) over 10 runs. Results show that \textbf{zero-shot} has the highest
variability, \textbf{one-shot} reduces it, and \textbf{two-shot + AQM} achieves the lowest
deviation. Among the LMs, \textbf{LLaMA3} is the most stable across prompting
strategies, suggesting that larger models are both more accurate and
more consistent than smaller ones such as \textbf{Phi-2}.

\section{Conclusion and Future Work}

This paper introduces a pipeline that translates high-level network intents into valid \texttt{tc} configuration commands through few-shot prompting combined with a semantic model. On a 100-intent benchmark, the AQM-guided two-shot prompting strategy with LLaMA3 attains 0.88 semantic similarity, 0.87 semantic unit coverage, and a normalized edit distance of 0.16, representing a 25–40\% gain over baseline prompting approaches. The sensitivity analysis across ten independent runs further confirms the robustness of our framework, showing low variability $(SD \leq 0.04)$ in performance metrics. 

The evaluation relies on a curated synthetic intent dataset, motivated by the lack of public benchmarks mapping declarative intents to validated \texttt{tc} configurations. While reflecting realistic operator objectives, it cannot fully capture real-world intent diversity. The semantic model is an offline, steady-state abstraction that enables deterministic grounding but does not capture transient dynamics. The proposed transformation targets Linux \texttt{tc} as a widely deployed mechanism for queueing and traffic shaping and is therefore scoped to policies expressible at this layer, with richer packet semantics or encrypted traffic requiring complementary higher-layer mechanisms. The pipeline operates at intent submission time with negligible critic overhead, and latency is dominated by language model inference for long-lived policies. Future work will address these aspects through telemetry-driven dynamic semantic models, iterative LM-critic refinement, and integration with orchestration platforms for production-level validation.

\bibliographystyle{IEEEtran}

\end{document}